\documentclass[usegraphicx,useAMS]{mn2e}
\usepackage{txfonts}

\newcommand\plotone[1]{\centering\includegraphics[width=\hsize]{#1}}

\newcommand\rEt{\tilde r_{\rm E}}
\newcommand\tE{t_{\rm E}}
\newcommand\tv{\tilde{v}}

\newcommand\dchi{\Delta\chi^2}

\def\change#1{{#1}}

\title
[The Nature of Parallax Microlensing Events Towards the Galactic Bulge]
{The Nature of Parallax Microlensing Events Towards the Galactic Bulge}

\author[Smith et al.]{
Martin C.~Smith,$^{1,2}$\thanks
{msmith@astro.rug.nl, vasily@ast.cam.ac.uk, 
nwe@ast.cam.ac.uk, smao@jb.man.ac.uk, jin@ast.cam.ac.uk}
Vasily Belokurov,$^3$\footnotemark[1]
N.~Wyn Evans,$^3$\footnotemark[1]
Shude Mao,$^2$\footnotemark[1]
Jin H.~An$^3$\footnotemark[1]\\
$^{1}$Kapteyn Institute, P.O. Box 800, 9700 AV Groningen, The Netherlands \\
$^{2}$Jodrell Bank Observatory, University of Manchester,
Macclesfield, Cheshire SK11 9DL, UK\\
$^{3}$Institute of Astronomy, University of Cambridge,
Madingley Road, Cambridge CB3 0HA, UK}

\date{
Accepted ........
Received .......;
in original form ......}

\pubyear{2004}

\begin{document}
\maketitle

\begin{abstract}
Perhaps as many as 30 parallax microlensing events are known, thanks
to the efforts of the MACHO, OGLE, EROS and MOA experiments monitoring
the bulge.  Using Galactic models, we construct mock catalogues of
microlensing light curves towards the bulge, allowing for the uneven
sampling and observational error bars of the OGLE-II experiment.
As a working definition of a parallax event, we require the
improvement $\Delta \chi^2$ on incorporating parallax effects in the
microlensing light curve to exceed 50. This enables us to carry out a
fair comparison between our theoretical predictions and
observations. The fraction of parallax events in the OGLE-II database
is around $\sim$1 per cent, though higher fractions are reported by
some other surveys. This is in accord with expectations from standard
Galactic models. The fraction of parallax events depends strongly on
the Einstein crossing time $t_{\rm E}$, being less than 5 per cent at
$t_{\rm E} \approx 50$ days but rising to 50 per cent at $t_{\rm E}
\ga 1$ yr.  We find that the existence of parallax signatures is
essentially controlled by the acceleration of the observer normalised
to the projected Einstein radius on the observer plane divided by
$\tE^2$.  The properties of the parallax events -- time-scales,
projected velocities, source and lens locations -- in our mock
catalogues are analysed.
Typically, $\sim38$ per cent of parallax events are caused by a disk
star microlensing a bulge source, while $\sim33$ per cent are caused
by a disk star microlensing a disk source (of these disk sources, one
sixth are at a distance of 5 kpc or less).
There is a significant shift in mean time-scale from 32 d for
all events to $\sim130$ d for our parallax events. There are
corresponding shifts for other parameters, such as the lens-source
velocity projected onto the observer plane ($\sim1110$ km~s$^{-1}$ for
all events versus $\sim80$ km~s$^{-1}$ for parallax events) and the
lens distance (6.7 kpc versus 3.7 kpc).
We also assess the
performance of parallax mass estimators and investigate whether our mock
catalogue can reproduce events with features similar to a number of
conjectured `black hole' lens candidates.
\end{abstract}


\section{Introduction}
\label{sec:intro}

Thousands of microlensing events in the Local Group have been
discovered by various collaborations, such as MACHO (e.g., Alcock et
al. 2000), OGLE (e.g., Wo\'zniak et al. 2001, Udalski 2003),
MOA (e.g., Sumi et al. 2003), EROS (e.g., Afonso et al. 2003) and
POINT-AGAPE (e.g., Paulin-Henriksson et al. 2003). The vast majority
of these are toward the Galactic centre, many of which were discovered
in real-time.\footnote{For example, see
http://www.astrouw.edu.pl/\~{}ogle/ogle3/ews/ews.html and
http://www.roe.ac.uk/\~{}iab/alert/alert.html} The database of
microlensing events provides a unique mass-selected sample to probe
the mass function of lenses and the mass distribution and dynamics of
the Galaxy (see Paczy\'nski 1996 and Evans 2003 for
reviews). Unfortunately, the mass of the lens can not be unambiguously
determined for most microlensing events because of
degeneracies. However, for the so-called exotic microlensing events,
which include the finite source size events (Witt \& Mao 1994; Gould
1994) and parallax microlensing events (Gould 1992), the degeneracies
are partly or wholly broken.  Astrometric microlensing offers another
exciting possibility to determine the lens mass (e.g., Walker
1995). Astrometric measurements may become feasible with the VLT
interferometer (Delplancke, G\'orski \& Richichi 2001) and future
satellite missions such as the {\it Space Interferometry Mission}
(Paczy\'nski 1998) and {\it GAIA} (Belokurov \& Evans 2002).

The standard microlensing light curve follows a characteristic
symmetric curve (e.g., Paczy\'nski 1986). However, this is based on
the assumption that the relative motions among the observer, lens and
source are all uniform and linear. This assumption is clearly wrong in
principle as we know the Earth revolves around the Sun, and
furthermore, the lens and source may be in binary systems of 
their own. However, as most microlensing events last of the order of
weeks, the effects of acceleration are not noticeable in most
events. Nevertheless, for some events, the resulting
departures from the standard curve are clearly visible. They
can range from a slight asymmetry to dramatic multiple peak behaviour
(Smith et al. 2002a). The significance of these so-called parallax
events is that they allow an additional constraint to be placed on the
lens mass. The parallax events are biased towards (i) more
massive, (ii) slow-moving or (iii) closer lenses. The first of these
biases means that parallax events offer a powerful way to detect
stellar remnants, such as neutron stars and stellar mass black holes
(Agol et al. 2003). Combined with other exotic effects, such as
finite source size effect, one can derive the lens mass uniquely
(e.g., Jiang et al. 2005).

\begin{table*}
\caption{Data for the known parallax events towards the Galactic
bulge. Two events (OGLE-1999-BUL-32/MACHO-99-BLG-22 and
MACHO-96-BLG-12/EROS-BLG-12) have two separate sets of best-fit
parameters since they have been observed and modelled by two different
collaborations. Errors are not given, but can be found in the relevant
reference. Values in parenthesis indicate the presence of two
degenerate parallax fits. Additional MACHO events have been presented
in Becker (2000), although the more convincing events from this sample
are presented below, since they also appeared in Bennett et
al. (2002a,b).}
\label{table:par}
\begin{tabular}{lcccl}
\hline
Lightcurve& 
$t_{\rm E}$ (d)&
$\tilde v$ (km s$^{-1}$)&
$\rEt$ (au)&
Reference\\
\hline
OGLE sc6\_2563	&  71.58 &	99.3 &	4.1  & Smith et al. (2002a) \\
OGLE sc20\_5748	&  78.2  &	53.9 &	2.4  & Smith et al. (2002a) \\
OGLE sc27\_3078	&  124   &	23.5 &	1.7  & Smith et al. (2002a) \\
OGLE sc33\_4505 &  194   &      56.9 &  6.4  & Smith et al. (2002a) \\
OGLE sc41\_3299	&  98    &	41.1 &	2.3  & Smith et al. (2002a) \\
OGLE sc43\_836	&  45.1  &	58.0 &	1.5  & Smith et al. (2002a) \\
OGLE sc26\_2218$^{\rm a,b}$&  39.53 &      167  &  3.8   & Smith, Mao,
Wo\'zniak (2003a)\\
OGLE-1999-BUL-19& 372.0	 &      12.5 &	2.7  & Smith et al. (2002b) \\
OGLE-1999-BUL-32& 640	 &      79   & 	29.1  & Mao et al. (2002) \\
- a.k.a. MACHO-99-BLG-22 & 560	 &      75  & 	24.3 & Bennett et al. (2002b) \\
OGLE-2000-BUL-43$^{\rm c}$& 156.4 (158.2)	& 40.1 (52.4) & 3.6 (4.8)& Soszy\'nski et al. (2002)\\
OGLE-2003-BLG-238$^{\rm a}$& 38.2 & 652.7 & 14.4 & Jiang et al. (2004) \\
OGLE-2003-BLG-175$^{\rm d}$& $\sim$63 & $\sim$141 ($\sim$106) & $\sim$5.1 ($\sim$3.9) & Ghosh et al. (2004)\\
MACHO-104-C	& 110	& 77	& 4.9 & Bennett et al. (2002a) \\
MACHO-96-BLG-5	& 485	& 30.9	& 8.7 & Bennett et al. (2002a)	\\
MACHO-96-BLG-12	& 147	& 47.5	& 4.0 & Bennett et al. (2002a) \\
 - a.k.a. EROS-BLG-12 & 145.6 & 43.7	& 3.7 & Afonso et al. (2003) \\
MACHO-98-BLG-6	& 245	& 79	& 11.2& Bennett et al. (2002a)	\\
MACHO-99-BLG-1	& 115.5	& 43.9	& 2.9 & Bennett et al. (2002a) \\
MACHO-99-BLG-8	& 120	& 62	& 4.3 & Bennett et al. (2002a) \\
EROS-BLG-29	& 108.3 & 69.8	& 4.4 & Afonso et al. (2003) \\
MOA-2000-BLG-11	& 69.7	& 42.5	& 1.7 & Bond et al. (2001) \\
MOA-2003-BLG-37$^{\rm d}$& $\sim$43 & $\sim$70 ($\sim$50) & $\sim$1.7 ($\sim$1.3) & Park et al. (2004) \\
\hline
\end{tabular}
\begin{minipage}{\hsize}
$^{\rm a}$This event also exhibits finite source signatures.\\ 
$^{\rm b}$\change{The parallax detection becomes marginal if the
possibility of small, negative blending is allowed in the fitting}.
\\
$^{\rm c}$Analysis of additional EROS data for this event has been
able to discriminate between the two degenerate sets of parameters,
showing that the fit with $t_{\rm E}=158.2~\mbox{d}$ is unfeasible (Le
Guillou 2003).\\
$^{\rm d}$Only approximate values are quoted as the parallax
parameters have been found to suffer from additional degeneracies.
\end{minipage}
\end{table*}

A systematic survey of parallax events of the MACHO database was
performed by Becker (2000). Bennett et al. (2002a) subsequently
published the most convincing long-duration events from this
survey. Smith, Mao \& Wo\'zniak (2002b) have searched systematically
for the parallax events in the 3-yr OGLE-II database. Parallax events
have also been found serendipitously in the MOA (Bond et al. 2001) and
EROS databases (Afonso et al. 2003). Table \ref{table:par} is
a compendium of the good and marginal parallax candidates in the
direction of the Galactic bulge. The fraction of parallax events in
the microlensing database ranges from around $\sim$1--10 per cent,
although some events are more convincing than others. 
Additional parallax events have been identified by Popowski et
al. (2004), although no model fits were presented for these candidates.
Important advances not listed in this table include the detection of the
parallax effect in the binary lens events EROS BLG-2000-5 (An et
al. 2002) and OGLE-2002-BLG-069 (Kubas et al. 2005), and in the Large
Magellanic Cloud single-lens event MACHO LMC-5 (Gould, Bennett \&
Alves 2004). For all of these events, the parallax effect leads to
an accurate determination of the mass of the lens, when combined with the
other astrometric or photometric data.

The question naturally arises whether the observed fraction is
consistent with theoretical expectations. This is an important
question as the number of parallax events depend on the mass function
and kinematics of lenses. There have been several previous studies
(Buchalter \& Kamionkowski 1997; Bennett et al. 2002a). However, these
studies have some deficiencies. For example, they all used regular
samplings and uniform simulated errors. Even sampling is
clearly a gross simplification as there are significant gaps in the
observational data, in particular, the annual period ($\sim$
late-October -- mid-February) during which the bulge cannot be
observed. This gap can be especially important for events of
sufficiently long duration, since the asymmetric nature of the
parallax signal will be more difficult to detect if a significant part
of the event lies within this gap.

The purpose of this paper is to conduct a study of parallax events
using Monte Carlo simulations for the OGLE-II experiment. We will
explicitly account for the uneven sampling, simulate realistic error
bars and adopt the same event selection criteria as Wo\'zniak et al.
(2001), in order to make a much better comparison between observations
and theoretical predictions.  The outline of the paper is as
follows. We first present the details of our simulations in
Section~\ref{sec:monte}. We then make mock catalogues of microlensing
events towards the Galactic bulge and compute the fractions of
observable parallax events in Section \ref{sec:freq}. We analyse the
properties of the parallax events in our mock catalogues in
Section~\ref{sec:obs}. {This section includes a number of subsections
dealing with event parameters (\ref{sec:tE} -- \ref{sec:rEdist}),
mass estimators (\ref{sec:estimators}) and long duration 
black-hole candidate events (\ref{sec:blackhole}).} We finish with a
summary and discussion of strategies for future parallax surveys in
Section \ref{sec:disc}.

\section{Monte Carlo Simulations}
\label{sec:monte}

\subsection{Parallax Microlensing Events} 

The observable quantities for parallax events are the Einstein radius
crossing time $t_{\rm E}$ and the projected velocity $\bmath{\tilde
v}$ on the observer plane. The former quantity is,
\begin{equation}
\label{eq:rEt}
t_{\rm E}=\frac{r_{\rm E}}{v_\perp}, \qquad\qquad
r_{\rm E}^2=\frac{4GM}{c^2} D_{\rm S} x(1-x),
\end{equation}
where $r_{\rm E}$ is the Einstein radius, $M$ is the mass of the lens,
$v_\perp$ is the speed of the lens transverse to the observer-source line of
sight, $D_{\rm S}$ is the source distance, $D_{\rm L}$ is the lens
distance, while $x=D_{\rm L}/D_{\rm S}$ is the ratio of the distance
of the lens to the source. The latter quantity is (Gould 1992),
\begin{equation}
\label{eq:vt}
\bmath{\tilde v}=  {\bmath{v_\perp} \over 1-x} = \frac{\bmath{v}_{\rm L} - x\bmath{v}_{\rm S}}{1-x}
-\bmath{v}_{\sun}
\end{equation}
where $\bmath{v}_{\rm L}$, $\bmath{v}_{\rm S}$ and $\bmath{v}_{\sun}$
are the velocities of the lens, source and the Sun transverse to the
line of sight. The Einstein radius crossing time $t_{\rm E}$ is always
measurable for well-sampled microlensing events; however, the
projected velocity is only measurable if the magnification
fluctuations caused by the motion of the Earth are substantial. The
observables ($t_{\rm E}, \bmath{\tilde v}$) can be used to construct
another useful quantity, namely the Einstein radius projected on the
observer plane
\begin{equation} 
{\tilde r_{\rm E}} = \tilde v t_{\rm E}.
\end{equation}

\subsection{The Galactic Model}

We assume that the sources and lenses may lie either in the Galactic
disk or the bulge. In practice, there may be some contamination from
sources in the Sagittarius dwarf galaxy (e.g., Evans 1995; Cseresnjes
\& Alard 2001). Bennett et al. (2002a) have argued that this may be
particularly important for the long duration events. Nonetheless,
the structure of the disrupting Sagittarius dwarf is too irregular and
uncertain for reliable modelling and so we do not include it as a
source population. In any case, Cseresnjes \& Alard argue that the
contribution from Sagittarius source events is less than 1 per cent
for the OGLE-II fields.

We adopt a value of 8.5 kpc for the Galactocentric distance of the
Sun.  Using standard cylindrical polars ($R,z$), the density law of
the Galactic disk is (e.g., Binney \& Evans 2001)
\begin{eqnarray}
\lefteqn{
\rho_{\rm L}(R,z)=
\rho_0\exp\left(-{R\over 3~\mbox{kpc}}\right)
}\nonumber\\&&\times
\left[0.435\,\mbox{sech}^2\left({z\over 220~\mbox{pc}}\right)
+0.565\exp\left(-{|z|\over 440~\mbox{pc}}\right)\right].
\label{eq:rhodisk}
\end{eqnarray}
This uses the vertical profile found by Zheng et al. (2001) in their
studies of disk M dwarfs with the {\it Hubble Space Telescope}. By
counting stars within 5 pc of the Sun (which can be detected through
their large proper motions) and using {\it Hipparcos} parallaxes
Jahrei{\ss} \& Wielen (1997) find that stars contribute
$3.9\times10^{-2}$ M$_{\sun}$~pc$^{-3}$ to the mass density at the
plane, which sets the local mass density and hence the overall
normalisation $\rho_0$. This is an accurate representation of the
local disk, embodying information from local star counts and stellar
kinematics. The velocity distribution of disk stars is taken as a
Gaussian with a mean $\langle v \rangle = (0, 214,
0)~\mbox{km~s$^{-1}$}$ in cylindrical polar coordinates. The disk
velocity dispersion is diagonalised along the same cylindrical polar
coordinate axes with $\sigma_{RR}=21~\mbox{km~s$^{-1}$}$,
$\sigma_{\phi\phi}=34~\mbox{km~s$^{-1}$}$ and
$\sigma_{zz}=18~\mbox{km~s$^{-1}$}$ (Edvardsson et al. 1993).  


The density law for the bulge deflectors is
%
\begin{eqnarray}\lefteqn{
\rho_{\rm L}(x',y',z)=}\nonumber\\&&
\rho_0\exp\left\{
-\left[\left({x'\over 990~\mbox{pc}}\right)^2
+\left({y'\over 385~\mbox{pc}}\right)^2
+\left({z\over 250~\mbox{pc}}\right)^2\right]^{1/2}\right\}
\label{eq:rhobulge}
\end{eqnarray}
%
with the major axis ($x'$-axis) in the Galactic plane and oriented at
$\sim$24\degr\ to the line of sight. This is the E2 model, as
suggested by Dwek et al. (1995) and subsequently modified by Stanek
(1997). It is a good fit to the near-infrared photometry of the bulge
as seen by the {\it COBE}. The normalisation $\rho_0$ is set to be 4.46 
M$_{\sun}$~pc$^{-3}$, which is obtained by setting the total mass
within 2.5 kpc of the Galactic centre to be $1.5\times10^{10}$
M$_{\sun}$.
The velocity distribution of the bulge
stars is a Gaussian about zero mean. The velocity dispersion tensor is
diagonal in the Cartesian coordinates along the axes of the triaxial
bulge with $\sigma_{x'x'}=114~\mbox{km~s$^{-1}$}$,
$\sigma_{y'y'}=86~\mbox{km~s$^{-1}$}$ and
$\sigma_{zz}=70~\mbox{km~s$^{-1}$}$ (c.f., Han \& Gould 1995; Evans \&
Belokurov 2000).

The mass function $\phi(M)$ -- or the number of stars per unit
mass -- is the multi-part power-law taken from equation (4) of Kroupa
(2002). This is an initial mass function and so coincides with the
present-day mass function below $\approx 1 M_\odot$. Above $1
M_\odot$, the index of the power-law is adjusted to give a good
representation of the luminosity function in the OGLE-II fields; this
gives a steep power-law of $-7$.  The luminosity function
corresponding to the mass function is computed in the following
way. Baraffe et al. (1998) provide mass-to-light coefficients in
different bands for stars of different ages with different chemical
abundances. We assume a stellar age of 5 Gyr and solar metallicity,
then compute the transformation from mass function to luminosity
function by numerical differentiation of this data. In our
simulations, masses are generated between $0.05 M_\odot$ and $5
M_\odot$. We do not generate masses above $5 M_\odot$, as the
probability distribution is a sharply decreasing function of mass. All
the stars with mass below $0.08 M_\odot$ are treated as dark.

\begin{figure*}
\plotone{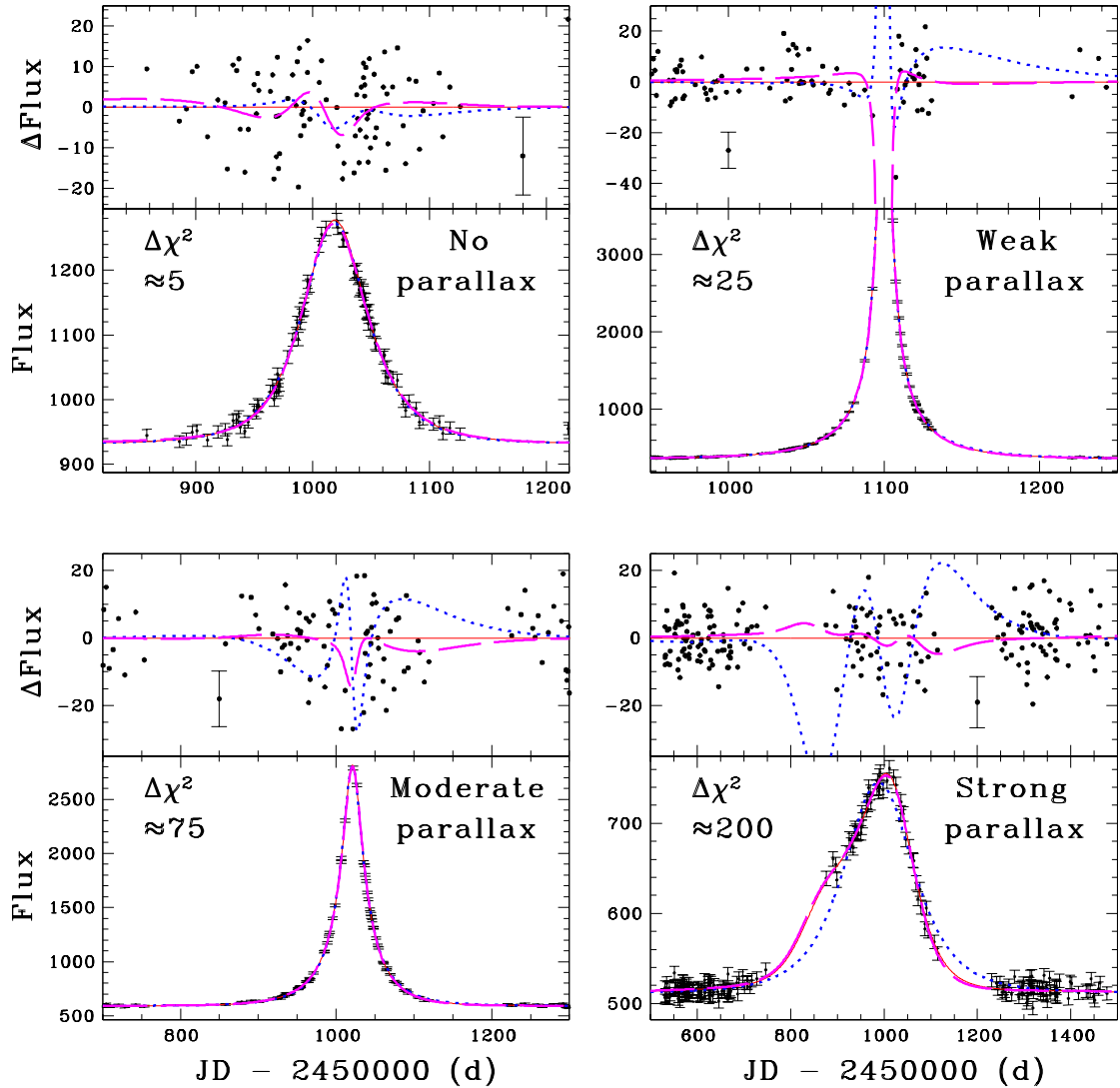}
\caption{Four sample light curves from our mock catalogues.  These
light curves illustrate the different cuts that are applied to the
catalogue to differentiate between no-, weak-, moderate-, and
strong-parallax signatures (top left, top right, bottom left and
bottom right, respectively). By our definitions, a no-parallax
event is one with $\dchi<10$. The solid line denotes the true light
curve, while the dotted and dashed lines correspond to the
best-fitting standard and parallax models, respectively. For each
light curve we also plot the residuals from the true light curve (top
panels). The error-bars have been omitted to aid clarity, although a
typical error bar has been included -- note that this is not an actual
data point from the event.}
\label{fig:mocklc}
\end{figure*}

\subsection{Simulation Algorithm}
\label{sec:simalg}

To simulate microlensing events, we must pick the event parameters
from the probability distribution
\begin{equation}
P(\ell, b, D_{\rm S}, m) \propto D_{\rm S}^2\ \rho_{\rm S}(\ell, b,
D_{\rm S})\Psi(m, \ell, b, D_{\rm S}) \Gamma (\ell, b, D_{\rm s})
\end{equation}
Here, $\rho_{\rm S}$ is the density of sources, $\Psi$ is the
luminosity function and $\Gamma$ is the microlensing rate at the
source location ($\ell, b, D_{\rm S}$). We ascribe the source
population to the bulge or disk according to the density at this
($\ell,b,D_{\rm S}$) and choose random velocity components according
to the source population. We note that the source velocity probability
distributions are separable.

The luminosity function $\Psi$ depends on position because of
extinction.  The $V$-band extinction at any location is calculated
using Drimmel \& Spergel's (2001) extinction law and translated into
$I$-band extinction using $A_V/E(V-I)=2.1$ (e.g., Popowski 2001). We
generate sources within the magnitude range $13.6<I< 21.0$. The lens
mass is generated from 
\begin{equation}
P(M) \propto M^{1/2}\phi(M).
\end{equation}
The flux contributed by the lens is calculated and the microlensing
event is retained only if $I <19$. The cut for objects fainter than
$I=19$ is applied to reduce the problem of blending by faint
background sources. Our algorithm therefore only takes the blending by
the lens light into account and applies a cut in magnitude to minimise
the effects of blending by faint background sources. Although, in
practise, the flux from the source can be contaminated by light from other
nearby stars, in Section \ref{sec:disc} we show that our results are
effectively unchanged if we incorporate a simple distribution for this
additional blended light.

The differential rate can be written out explicitly as
\begin{equation}
{\rm{d}^4\Gamma\over \rm{d}^2{\bmath{v}_{\rm L}} \rm{d}D_{\rm L} \rm{d}M} = {\phi(M)\over
\langle M \rangle}
\rho_{\rm L}(\ell,b,D_{\rm L}) F(\bmath{v_\perp}) 2v_\perp
r_{\rm E}(M,D_{\rm L})
\end{equation}
where $\rho_{\rm L}$ is the total density of lenses,
$F(\bmath{v_\perp})$ is the distribution of relative transverse lens
velocities, $\langle M \rangle$ is the average mass while $\bmath{v_\perp}$
is the relative transverse velocity of the lens, as defined in
equation~(\ref{eq:vt}). So, for the lenses, we choose the distance $D_{\rm
L}$ and velocity $\bmath{v}_{\rm L}$ from the probability
distribution:
\begin{equation}
P(D_{\rm L}, \bmath{v}_{\rm L}) \propto \sqrt{ D_{\rm L} (D_{\rm
S}-D_{\rm L})} \rho_{\rm L}(D_{\rm L}) F(\bmath{v_\perp}) v_\perp
\end{equation}
The lens population is chosen in this step by comparing the relative
densities of disk and bulge lenses at this location.  Finally, an
impact parameter (defined as the minimum separation between the lens
and the Sun-source line of sight, in units of the Einstein radius) is
generated in the range $\in[-2,2]$.
This approach to selecting the impact parameter means than a fraction
of events will be incorrectly omitted for $\rEt\la1$, since the peak
magnification depends on the separation between the lens and the
Earth-source line of sight, not the Sun-source line of sight; however,
this problem is negligible since only 0.3 per cent of events from
our model have $\rEt<1$.

This prescription gives us a microlensing event for our mock catalogue.

\subsection{Mock Light Curves and Catalogues} 
\label{sec:mocklc}

The characteristics of the light curves were chosen to match the 520
event OGLE-II Difference Image Analysis (DIA) microlensing catalogue
of Wo\'zniak et al. (2001; see also Wo\'zniak (2000) for a detailed
description of the OGLE-II data). These OGLE-II catalogues are
constructed from three years of Galactic bulge observations, with
observations taken in the $I$-band once every few nights during the bulge
season (typically mid-February until the end of October) resulting in
between 200 and 300 observations per light curve. The OGLE-II
experiment consists of 49 bulge fields covering a total area of
approximately 10 deg$^2$ (see fig.~1 of Wo\'zniak et al. 2001). Two of
the fields are monitored much less frequently and a further three
fields had no observations during the first season. The limiting
magnitude of the experiment is $I\approx20$ and the saturation limit
is $I\approx11.5$.

To obtain the time sequence of observations we randomly select 100
OGLE-II light curves for each field from the variable star light
curves of Wo\'zniak et al. (2002), and then pick one of these
time-series at random for each simulated event in the field. We then
calculate the flux for each epoch. To do this, we assume that the
photometric errors are Gaussian and scale according to the following
empirical relation derived from the variable light curves of
Wo\'zniak et al. (2002),
\begin{equation}
\sigma_F = 5.04 + 3.58\times10^{-3}\ F^{1.04}.
\end{equation}
Here, the flux ($F$) and the error on the flux ($\sigma_F$) are in
units of 10 ADU. These fluxes can be converted into $I$-band
magnitudes through,
\begin{equation}
I(t) = 23.35 - {\rm log}~F(t).
\end{equation}
As stated in Section \ref{sec:simalg}, we simplify our
simulations by incorporating the blended flux from the lensed star
only, i.e. it is assumed that all of the observed flux comes from 
either the lensed source or the lens and not from any other nearby
stars.
  
As a final step, we implement the microlensing detection
criteria of Wo\'zniak et al. (2001). These criteria were employed to
discriminate against variable stars in the OGLE-II DIA catalogue and
must be applied to our mock catalogues in order to select only events
with noticeable brightening and a sufficiently constant baseline.

\begin{figure}
\centering\includegraphics[height=9cm]{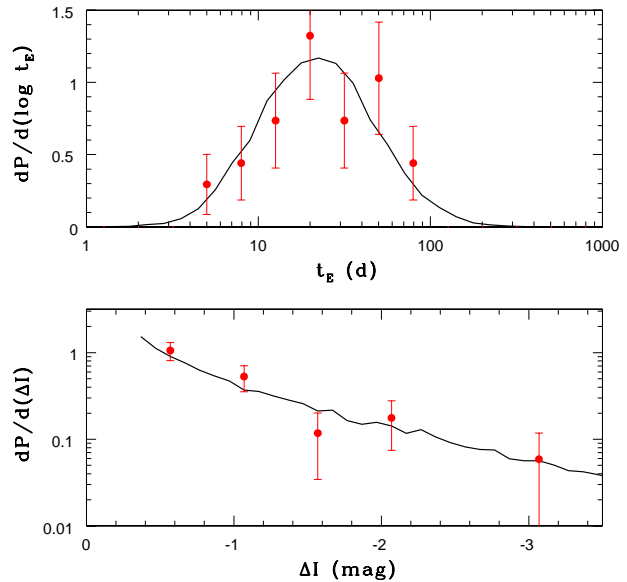}
\caption{Comparison between our mock catalogue (solid line) and the
observed OGLE-II catalogue of 33 bright events (data points) from Sumi
et al. (2005), where we have assumed Poisson errors. The top panel
shows the time-scale distribution and the lower panel shows the
amplitude of the magnification at the peak of the event. To make a
fair comparison with the catalogue of Sumi et al (2005) we have only
plotted events from our mock catalogue that are unblended, have
magnifications greater than $3/\!\sqrt{5}$ and have baseline magnitude
brighter than $I=17$.}
\label{fig:checkcat}
\end{figure}


To assess the credibility of our mock catalogue, we compare it to
the observed OGLE-II catalogue of 33 bright events from Sumi et
al. (2005). This catalogue of Sumi et al. (2005) is based on 
events that lie within an extended Red Clump Giant region and are
selected to be unblended. We choose this catalogue rather than the 520
event catalogue of Wo\'zniak et al. (2001) as no detailed modelling
was carried out on this larger sample. In addition, the catalogue of
Wo\'zniak et al. (2001) is affected by blending from unrelated stars
near to the source, whereas our mock catalogues only consider blending
from the lens.
In Fig.~\ref{fig:checkcat} we compare two
properties, the $\tE$ distribution and the distribution of
amplification at the peak. As can be seen from this figure, the two
catalogues are in good agreement. For example, the values of the mean
$\tE$ are 29.4 d and 28.9 d for the mock catalogue and the observed
catalogue, respectively.

\begin{table}
\begin{tabular}{lccc}
\hline
S/N cut & \multicolumn{3}{c}{Percentage of parallax events} \\
\null & Weak parallax & Moderate parallax & Strong parallax \\
& ($\Delta\chi^2>10$) & ($\Delta\chi^2>50$) & ($\Delta\chi^2>100$) \\
\hline
5 & 4.6 & 1.4 & 0.8 \\
15 & 7.1 & 2.3 & 1.5 \\
30 & 10.0 & 3.6 & 2.3 \\
\hline
\end{tabular}
\caption{The percentage of parallax events using the criteria defined in
Section~\ref{sec:def} as a function of S/N cut.}
\label{table:paratot}
\end{table}

\section{Frequency of Parallax Events}
\label{sec:freq}

\subsection{Working Definition of A Parallax Event}
\label{sec:def}

As a first step in analysing our mock catalogues, we fit all events
with both the five-parameter blended Paczy\'nski light curve and
seven-parameter blended parallax model. The improvement afforded by
the parallax model is recorded as the improvement in
$\chi^2$.
We also calculate a measure of the signal-to-noise ratio (S/N) for
each light curve, defined as the minimum value of $F/\sigma_F$ for the
three points bracketing the maximum amplification.

Before comparing theoretical predictions with observations, we must
first investigate what level of improvement in $\chi^2$ is required to
classify an event as a parallax event. This is important because it is
conceivable that problems such as scatter in the data could be
misidentified as parallax signatures.

\begin{figure}
\plotone{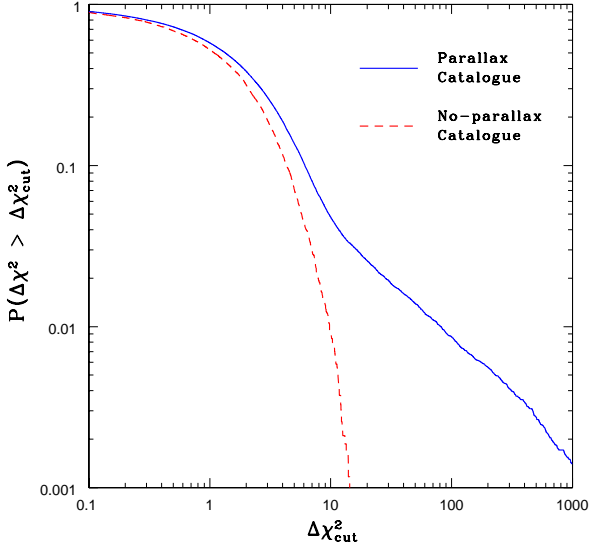}
\caption{The distribution of $\Delta\chi^2$ for our simulated
catalogue (solid) and for the same catalogue but with no parallax
signatures (dashed). All events have $\mbox{S/N}>5$.}
\label{fig:nopara}
\end{figure}

To test this, we construct a catalogue of events using our
Galactic model but generate light curves with no parallax
signatures. We then fit these events with both the standard and
parallax models and investigate the distribution of
$\dchi=\chi^2_{\rm stan}-\chi^2_{\rm para}$. Fig. \ref{fig:nopara} 
shows the distribution of $\Delta\chi^2$ for both the parallax and
no-parallax catalogues. The no-parallax distribution shows a rapid
decline in the fraction of events with $\dchi>10$ (less than 1 per
cent have $\dchi>10$). This rapid decline indicates the minimum cut
that should be applied in order to identify parallax events.
However, when dealing with real data, we must also bear in mind that
there may be contamination from other effects, such as binary
signatures and/or problems with the data.

Throughout this paper, we adopt three different designations for
parallax events: convincing or strong events, with $\dchi>100$;
moderate events, with $\dchi>50$; and marginal or weak events, with
$\dchi>10$. (If one uses the $F$-test for the significance of
parameters [see Smith et al. 2002b], these limits correspond to ${\rm
log}~p_F < -2,-10,-20$, respectively). Unless otherwise stated, when
we refer to parallax events, the `moderate' criterion is implied,
i.e. $\dchi>50$. In addition, unless otherwise stated, we restrict
ourselves to good quality parallax events with $\mbox{S/N}>5$ (as
defined at the beginning of this section). In Fig.~\ref{fig:mocklc},
we show four sample light curves from our mock catalogues. These light
curves show clearly how the standard model becomes increasingly less
able to fit the mock data as $\dchi$ becomes greater. They also
illustrate that gaps in the data (in particular, the three month gap
between observing seasons) significantly affect our ability to
identify parallax events.

\begin{figure}
\plotone{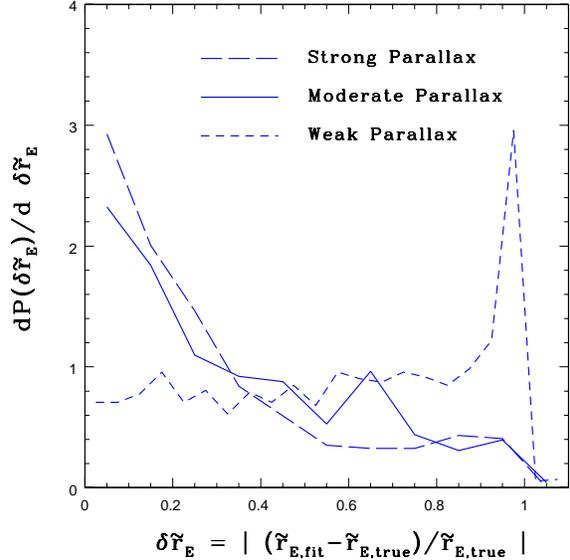}
\caption{The distribution of fractional error in the recovered
projected Einstein radius, i.e. 
$\delta \tilde{r}_{\rm E} = |(\tilde{r}_{\rm E,fit}-\tilde{r}_{\rm E,true})/\tilde{r}_{\rm E,true}|$.
This is shown for our three levels of parallax events, strong (long
dash), moderate (solid line) and weak (short dash). The large
fraction of weak parallax events with $\delta \tilde{r}_{\rm E}
\approx 1$ are caused by events with $\tilde{r}_{\rm E,fit} \approx 0$.}
\label{fig:paraerr}
\end{figure}

The issue of classifying parallax events is further complicated by the
fact that even though an event may be displaying parallax signatures,
this does not necessarily imply that the parallax parameters can be
recovered with a high degree of accuracy due to the degeneracies that
are inherent in the parallax formalism (see Gould 2004 and references
therein for details of the various types of degeneracies -- both
continuous and discrete -- that can affect parallax events). We
illustrate this in Fig. \ref{fig:paraerr} by investigating the
distribution of fractional error in recovered projected Einstein radius,
i.e. $\delta \tilde{r}_{\rm E}=|(\tilde{r}_{\rm E,fit}-\tilde{r}_{\rm E,true})/\tilde{r}_{\rm E,true}|$.
We find that the fraction of events with
$\delta \tilde{r}_{\rm E}<0.3$ is 64, 53 and 23 per cent for our
strong, moderate and weak parallax events, respectively. The large
fraction of weak parallax events with $\delta {\tilde{r}_{\rm E}}
\approx 1$ are caused by events with $\tilde{r}_{\rm E,fit} \approx 0$. A
comprehensive investigation into the nature of the errors for the
fitted parameter $\rEt$ is beyond the scope of this paper.

\subsection{Fraction of Parallax Events}
\label{sec:parafrac}

\begin{figure}
\plotone{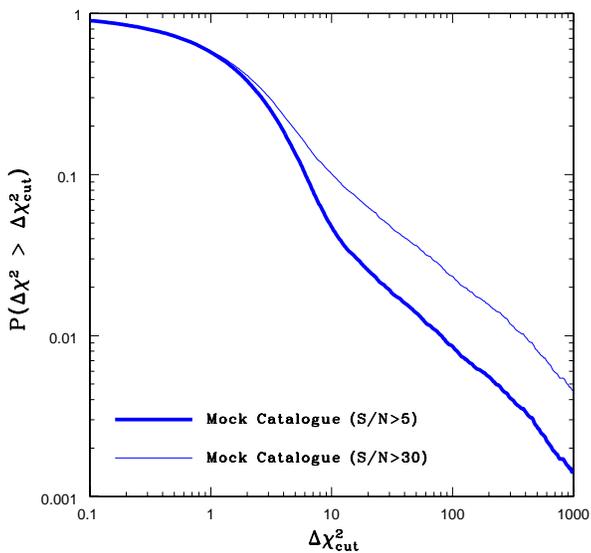}
\caption{The cumulative fraction of parallax events in the mock
catalogue, i.e. the fraction of parallax events for
different cuts of $\Delta\chi^2$. The thick and thin lines 
are for events with $\mbox{S/N}>5$ and $\mbox{S/N}>30$
respectively.}
\label{fig:paratot}
\end{figure}

We now analyse the mock catalogue in order to calculate predicted
fractions of parallax events. To do this, we compute the total number
of events that have $\dchi$ greater than a given value. These
fractions are plotted in Fig.~\ref{fig:paratot} as a function of the
$\Delta\chi^2$ cut for two different choices of S/N cut. This figure
shows that for moderate parallax events (i.e. $\Delta\chi^2>50$) our
catalogues predict a total of $\sim$1.4 per cent from all events with
$\mbox{S/N}>5$ or $\sim$3.6 per cent for $\mbox{S/N}>30$. We tabulate
our findings in Table~\ref{table:paratot}.
For $\Delta\chi^2_{\rm cut}>10$, the slope of these relationships can be
approximated by the following power-laws,
\begin{eqnarray}
\lefteqn{
P(\Delta\chi^2>\Delta\chi^2_{\rm cut})=0.22(\Delta\chi^2_{\rm
  cut})^{-0.71} ~{\rm for}~(\Delta\chi^2_{\rm cut}>10, {\rm S/N}>5)}\\
\lefteqn{
P(\Delta\chi^2>\Delta\chi^2_{\rm cut})=0.44(\Delta\chi^2_{\rm
  cut})^{-0.64} ~{\rm for}~(\Delta\chi^2_{\rm cut}>10, {\rm S/N}>30).}
\end{eqnarray}

We have also computed the
fraction of parallax events on replacing our standard Galactic disk
model with a maximal disk model, in which the local disk surface
density is increased to 100 $M_\odot$ pc$^{-2}$. The rationale for
this is that local disk stars are good candidates as lenses for
parallax events, and so the fraction may be a good diagnostic of the
local disk density. However, while this change does produce
slightly more parallax events, the effect is small and dwarfed by
the observational uncertainties.

The total fractions from our mock catalogues are
in broad agreement with a previous study by Buchalter \&
Kamionkowski (1997), which estimated that $\sim$1 per cent of
microlensing events towards the bulge should exhibit noticeable
parallax signatures.  The results from Fig.~\ref{fig:paratot} are also
in rough agreement with the fraction of parallax events from different
microlensing collaborations, such as MACHO, OGLE and MOA. From a
catalogue of 321 events from a 7-yr survey by the MACHO collaboration,
approximately $\sim$2 per cent exhibited convincing parallax
signatures and approximately $\sim$3 per cent exhibited weak parallax
signatures (Bennett et al.  2002a). The MOA collaboration found one
parallax events (Bond et al.  2001) in their 20-event catalogue from
bulge observations in the year 2000, i.e. $\sim$5 per cent. The
fraction of parallax events in the OGLE-II catalogues can be found in
two studies: Smith et al. (2002b) found 1 convincing parallax event
from a sample of 512 events, although a number of parallax events were
subsequently found to have been omitted from this 512 event sample
(e.g. Mao et al. 2002; Smith et al. 2002a); and Sumi et al. (2005)
found 1 parallax event~\footnote{This parallax event is the same as the
one discovered in Smith et al. 2002b.} in their sample of $\sim30$
bright red clump giant sources.
Although the above results seem to be consistent, a recent study by
the EROS collaboration (Afonso et al. 2003) reported a much higher
fraction of parallax events; from a total of 16 red clump giant events
they identified 2 convincing parallax events, i.e. 12.5 per cent. We
return to the issue of these EROS parallax events in the
discussion. However, the level of discrepancy (or agreement) between
these fractions must be taken with caution as all are subject to small
number statistics.  In addition, such comparisons are affected by the
different properties of the catalogues, i.e. the observing strategy
and duration of each project, etc. When comparing such fractions, we
should also bear in mind the fact that our mock catalogues are
restricted to events with $I$-band baseline magnitude brighter than 19
mag.


\begin{figure}
\plotone{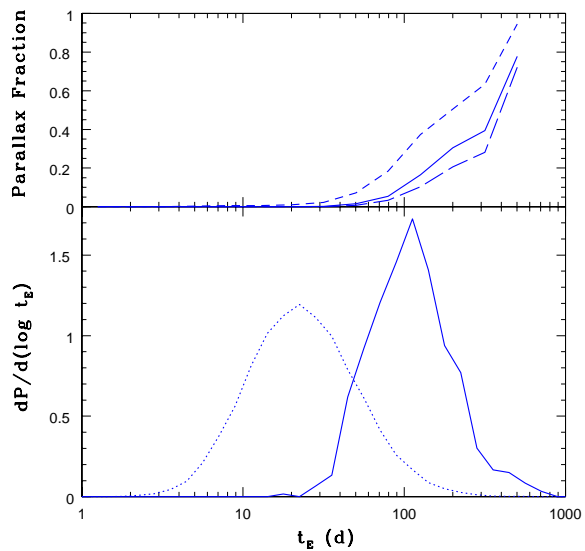}
\caption{ The lower panel shows the distribution of event time-scales
for all events (dotted line) and parallax events (solid line) in our
mock catalogue. The upper panel shows the corresponding fraction
of parallax events as a function of time-scale for strong,
moderate and weak parallax events (long dash, solid and short dash,
respectively).  Note that these plots use the true event time-scale,
not the fitted event time-scale.}
\label{fig:paratE}
\end{figure}

\begin{table*}
\begin{tabular}{lcccc}
\hline
& \multicolumn{4}{c}{Mean (median) values} \\
& All events & Weak parallax & Moderate parallax & Strong
parallax \\
\hline
$t_{\rm E}$ (d) & 32.1 (22.9) & 91.2 (73.4) & 130.4 (105.8) & 141.6 (112.0)\\
$\rEt$ (au) & 10.3 (8.3) & 6.4 (5.2) & 4.8 (3.9) & 4.5 (3.8)\\
$\tilde v$ (km~s$^{-1}$) & 1108.6 (609.0) & 247.2 (113.9) & 80.1 (61.4) & 69.6 (52.8) \\
$D_{\rm S}$ (kpc) & 9.7 (9.2) & 9.6 (9.0) & 9.2 (8.6) & 9.1 (8.6) \\
$D_{\rm L}$ (kpc) & 6.7 (7.0) & 4.7 (4.4) & 3.7 (3.1) & 3.5 (2.9) \\
\hline
\end{tabular}
\caption{Mean (median) values for the microlensing parameters
discussed in Section~\ref{sec:obs}.}
\label{table:paramean}
\end{table*}

\begin{figure*}
{\centering\includegraphics[width=10cm]{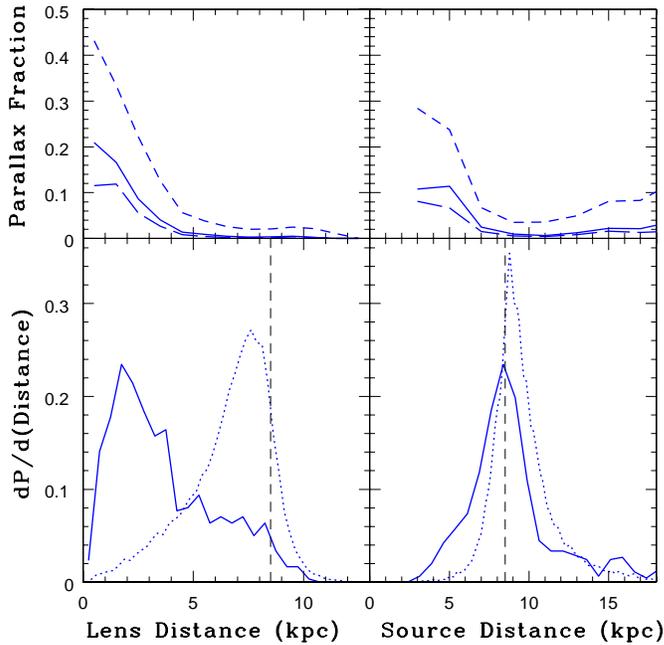}}
\caption{The upper panels show the fraction of events that are
classified as parallax as a function of lens/source distance for
strong, moderate and weak parallax events (long dash, solid and short
dash, respectively).  The lower panels show the distributions of lens
and source distances for all events (dotted line) and parallax events
(solid line) in our mock catalogue.  The vertical dashed lines at 8.5
kpc denote the centre of the Galaxy in our simulations.}
\label{fig:paraD}
\end{figure*}

\section{The Properties of the Parallax Events}
\label{sec:obs}

In this section, we investigate various properties of parallax events
in our mock catalogue in order to understand their nature. We utilise
the criteria defined in Section~\ref{sec:def} to refer to strong,
moderate and weak parallax events and enforce a minimum S/N of 5. We
deal with the event timescales, lens and source distances, and
projected velocities in turn. Table \ref{table:paramean} summarises
the mean and median values for these quantities. We also assess the
performance of parallax mass estimators and investigate whether our mock
catalogue can reproduce events with features similar to a number of
conjectured `black hole' candidates.

\subsection{Event Time-scale}
\label{sec:tE}

Fig.~\ref{fig:paratE} shows the distribution of the true event
time-scale~\footnote{It should be noted that the underlying
distribution of time-scale is mainly controlled by the choice of mass
function (see, for example, the left hand panel of figure 10 in Alcock
et al. 2000); however, the purpose of our work is to investigate
parallax events and so we do not concern ourselves with undertaking a
detailed analysis of the overall form of the time-scale distribution.}
(i.e. not the fitted time-scale). As expected, the parallax events all
lie in the long duration tail of the time-scale distribution. While
the $t_{\rm E}$ distribution for all events in our mock catalogue
peaks around $t_{\rm E}\approx20~\mbox{d}$, parallax events peak
around $t_{\rm E}\approx 100~\mbox{d}$. Although we might expect
parallax events to be detected at very large time-scales, few are seen
with $t_{\rm E}\ga 500~\mbox{d}$. This is because our simulated light
curves are based on the 3-yr dataset from OGLE-II. The microlensing
event detection drops off sharply for $t_{\rm E}\ga500~\mbox{d}$, as
such long events do not pass the constant baseline criterion necessary
to differentiate microlensing events from variable stars (Wo\'zniak et
al. 2001).

Given the quality and sampling of the OGLE-II data, it is unlikely
that we can detect parallax signatures for events with $t_{\rm
 E}\la50~\mbox{d}$, even for weak parallax events. The upper panel
of Fig. \ref{fig:paratE} shows that at $t_{\rm E}\approx50~\mbox{d}$,
the fraction of events that display parallax signatures is less than $\sim5$
per cent (0.7, 1.5, and 7.2 per cent for strong, moderate and weak
parallax events). The prospect of obtaining or constraining the
parallax parameters for such short duration events can be helped by
the more-frequent sampling and/or improved photometric accuracy, such
as that afforded by the numerous microlensing follow-up networks
(e.g. Jiang et al. 2005); however, it is known that such short
duration events are affected by significant degeneracies (e.g. Gould
1998).

The fraction of events that display moderate parallax signatures
exceeds 50 per cent for events with $t_{\rm E}\ga1~\mbox{yr}$.  However, the fact
that the fraction of parallax events only reaches $\sim$50 per cent
for $t_{\rm E}\approx1~\mbox{yr}$ shows that even for such long
duration events, the presence of strong parallax signatures is not
guaranteed.

\begin{figure*}
{\centering\includegraphics[width=10cm]{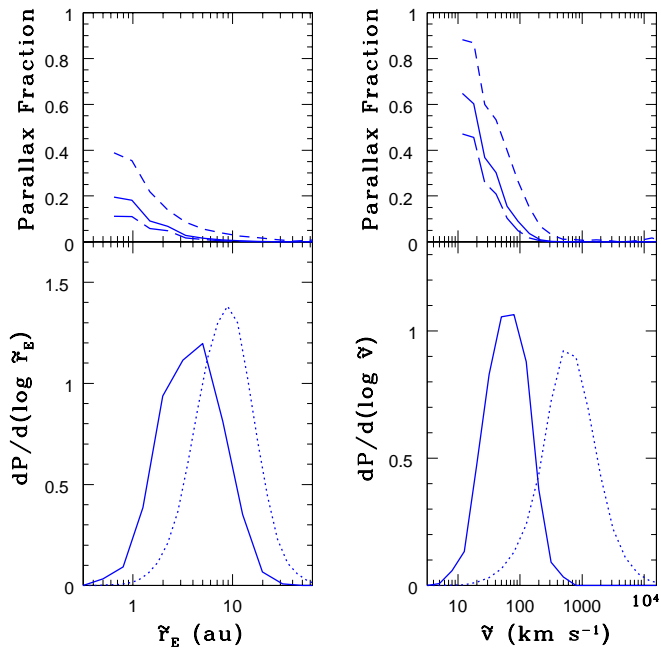}}
\caption{The lower panels show the distribution of the Einstein
radius projected into the observer plane ($\rEt$; left) and the
projected lens velocity on the observer plane ($\tilde v$; right) for
all events (dotted line) and parallax events (solid line).  The upper
panels show the fraction of events that display parallax
signatures as a function of $\rEt$ (left) and $\tilde v$ (right) for
strong, moderate and weak parallax events (long dash, solid, and short
dash, respectively).
Note that these plots use the true values for $\rEt$
and $\tilde v$, not the fitted values.}
\label{fig:pararEt}
\end{figure*}
\begin{figure}
\plotone{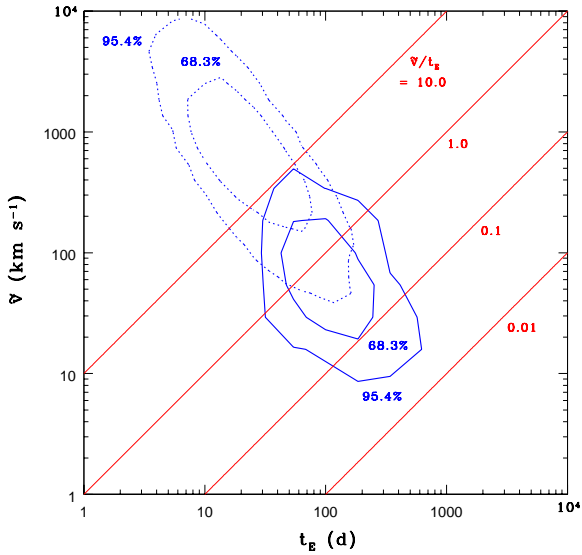}
\caption{Contour plot showing the joint distribution of event
time-scale and projected lens velocity on the observer plane for all
events (dotted) and parallax events (solid). The four diagonal lines
indicate constant $\tv / \tE = \rEt/\tE^2 =
10.0,1,0.1,0.01~\mbox{km~s$^{-1}$~day$^{-1}$}$.
If we approximate the Earth's projected acceleration to be
$a \approx 0.3~\mbox{km~s$^{-1}$~day$^{-1}$}$ (i.e. averaged over the whole
year), then these contours correspond to ${\cal A} = a/(\tv /
\tE)\approx 0.03,0.3,3,30$.
This dimensionless acceleration parameter ${\cal A}$
effectively controls the significance of the parallax deviations (see
Section \ref{sec:rEdist}).}
\label{fig:vtEcont}
\end{figure}

\subsection{Lens and Source Distances}
\label{sec:obs:dist}

Fig.~\ref{fig:paraD} shows the distributions of lens and source
distances. The lensed sources are preferentially on the far side of
the Galactic centre (see also Table~\ref{table:paramean}, which shows
that the mean source distance for all events is 9.7 kpc), as these
sources have more foreground lensing stars, i.e. the optical depth is
greater for sources behind the Galactic centre. They will appear
fainter (Stanek 1995) and may have different proper motions from the
overall population of stars (Mao \& Paczy\'nski 2002).

From Fig.~\ref{fig:paraD}, it can be seen that nearby lenses are
favoured for parallax events. This is due to the fact that nearby
lenses have smaller values of $\tv$ and $\rEt$ owing to the
smaller projection factor (see Section \ref{sec:rEdist}).

It has been suggested (e.g., Smith et al. 2002b) that parallax events
may be preferentially caused by 
so-called disk-disk lensing, which refers to events in which both the
source and lens lie in the disk.  For such events the lens, source and
observer are all co-rotating with the Galactic disk, producing small
relative transverse velocities and hence enhancing the probability of
detecting parallax signatures. This phenomenon can be seen in the
right-hand panel of Fig.~\ref{fig:paraD}, which exhibits a peak in the
fraction of parallax events for nearby sources. However, the
ratio of parallax events with disk sources is reduced due to the fact
that the overall density of sources in the near side of the disk is only
small. For our Galactic model we predict that $\sim$5 per cent of
parallax events have lensed sources lying within 5 kpc, which is a
significant fraction when one considers that only 0.5 per cent of all
microlensing events have sources located in this region.

\begin{table*}
\begin{tabular}{lccccc}
\hline
Lens & Source & \multicolumn{4}{c}{Percentage of events with given
  lens-source configuration} \\
\null & \null & All events & Weak parallax & Moderate parallax & Strong
parallax \\
\hline
near disk & near disk &
3.1 (0.5) & 19.4 (3.1) & 33.2 (5.2) & 36.3 (5.7) \\
near disk & far disk &
12.0 (0.9) & 9.4 (3.4) & 7.5 (4.7) & 8.9 (5.1) \\
far disk  & far disk & 
1.5 (0.1) & 0.9 (0.0) & 0.5 (0.0) & 0.8 (0.0) \\
disk & bulge&
29.6 (17.1) & 39.6 (49.7) & 38.4 (62.8) & 36.9 (65.9) \\
bulge & bulge&
34.0 (71.6) & 14.3 (31.3) & 8.9 (17.4) & 7.3 (13.3) \\
bulge & disk&
19.9 (9.8) & 16.3 (12.4) & 11.6 (9.9) & 9.8 (10.0) \\
\hline
\end{tabular}
\caption{
The percentage of events as a function of lens-source configuration for
all events and weak-, moderate-, and strong-parallax events.
For the values without parentheses, a disk star is
defined as one drawn from the population described by equation
(\ref{eq:rhodisk}), while a bulge star is drawn from
equation~(\ref{eq:rhobulge}). Bennett et al. (2002a) used a different
nomenclature, defining all stars within 3.5 kpc of the Galactic centre
as ``bulge'' and all stars outside 3.5 kpc of the Galactic centre as
``disk''. To enable comparison with Bennett et al.'s results, we give
in parentheses the values corresponding to these definitions. Note
that the disk-disk events are subdivided according to whether the disk
stars reside on the near- and far-side of the Galactic centre.}
\label{table:parafrac}
\end{table*}

In Table \ref{table:parafrac}, we compare the relative abundance of
parallax events within the different lens-source configurations, such
as for disk-bulge events (i.e. events in which the lens is in the disk
and the source is in the bulge), etc. Clearly, one can see
that the majority of parallax events are due to the disk-disk and
disk-bulge events, namely $\sim33$ and $\sim38$ per cent
respectively. This can be contrasted with the most common
configuration for all microlensing events, which is bulge-bulge.
Note that our parallax events have $\sim5$ per cent of sources within
5 kpc, which is slightly larger than the estimate of less than 3 per
cent that was predicted by the simulations of Bennett et al. (2002a).

\subsection{Projected Einstein Radius and Projected Lens Velocity}
\label{sec:rEdist}

Fig.~\ref{fig:pararEt} plots the distribution of the Einstein radius
projected into the observer plane $\rEt$ and the projected velocity
$\tilde v$ for all events and for parallax events. This shows that
parallax signatures are more readily detectable for events with small
values of $\rEt$ and $\tilde v$, as predicted by Smith et
al. (2002a). This can be understood, on considering that the projected
velocity of the lens should be comparable to (and preferably less than)
the orbital velocity of the Earth, which is
$v_\oplus\sim30~\mbox{km~s$^{-1}$}$.  The reason why smaller values of
$\rEt$ are favoured is because this parameter determines the length
scale on which the magnification is calculated. This means that the
magnitude of the deviations is determined by the magnitude of the
Earth's motion relative to this projected Einstein radius, i.e. in
general, the larger the projected Einstein radius, the smaller the
deviations. In Fig.~\ref{fig:vtEcont}, we show the joint distribution
of $t_{\rm E}$ and $\tilde v$, which -- as expected -- shows that
parallax events preferentially have large $t_{\rm E}$ and small
$\tilde v$. From this figure, it is clear that the ability to detect
parallax signatures is controlled by the quantity $\tv / \tE = \rEt /
\tE^2$.  For a given acceleration $a$, the dimensionless parameter
${\cal A}= a/(\rEt / \tE^2)$ describes the deviation of the trajectory
from a straight line (Smith, Mao \& Paczy\'nski 2003b).  Parallax
deviations are caused by the Earth's orbital acceleration. Although
the magnitude of this acceleration is constant in the
ecliptic plane, when projected into the lens plane it is no
longer constant and so $a$ varies throughout the year. However, most
parallax events last for a sufficiently long duration for the average
value of $a$ to be used. This means that the quantity $\rEt/\tE^2$
effectively determines the value of ${\cal A}$ and hence controls the
significance of the deviation due to the Earth's motion, i.e. a larger
$\rEt/\tE^2$ indicates that the parallax effect will be more difficult
to detect and vice versa.  We return to the issue of the $\rEt$
distribution in the discussion, where we compare our predictions to the
observed distribution.


\begin{figure*}
{\centering\includegraphics[width=10cm]{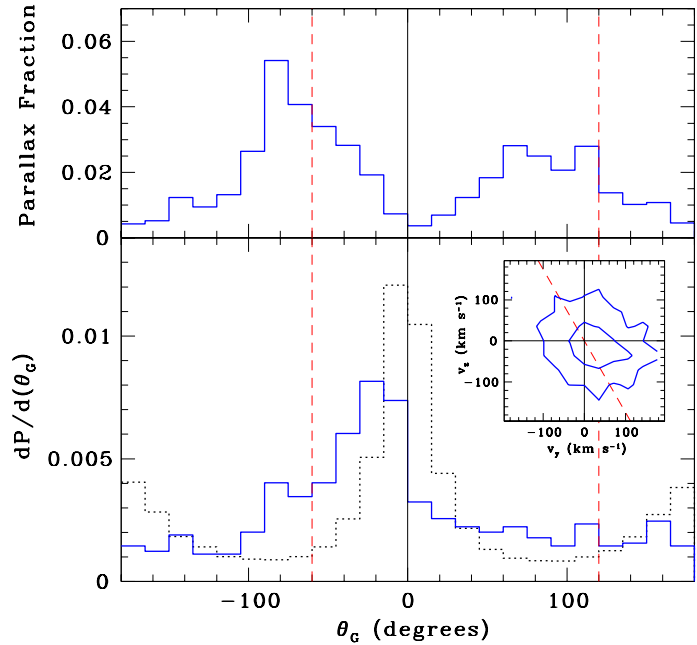}}
\caption{The lower panel shows the distribution of the
projected velocity angle, $\theta_G$, for all events (dotted
line) and moderate parallax events (solid line) in our mock
catalogue. $\theta_G$ is measured in Galactic coordinates from the
Galactic plane (in the direction of rotation) towards the North
Galactic pole.  The inset shows the projected velocity
distribution in Galactic coordinates, where the horizontal axis
corresponds to the direction of rotation and the vertical axis
corresponds to the North Galactic Pole. The contours
correspond to the 68.3\% and 95.4\% confidence intervals.
The upper panels show the fraction of events that are classified as
moderate parallax events as a function of the projected velocity
angle.  The dashed lines denote the orientation of the ecliptic plane
in this coordinate system.}
\label{fig:angles}
\end{figure*}

For a parallax event, it is possible to determine the orientation of
$\mathbf{\tilde{v}}$, from which we can gain additional information
regarding the event. In Fig. \ref{fig:angles}, we show the
distribution of this angle transformed into Galactic coordinates
($\theta_G$), where $\theta_G$ is measured from the Galactic plane (in
the direction of rotation) towards the North Galactic pole. From this
figure, it can be seen that parallax events preferentially have
$\mathbf{\tilde{v}}$ orientated in a direction parallel to the
ecliptic plane.  This can be understood when one considers that
trajectories perpendicular to the ecliptic plane will have less time
to be affected by the Earth's motion, compared to trajectories that
pass parallel to the ecliptic plane, i.e. for perpendicular
trajectories the Earth's acceleration will only affect the peak of the
light curve (Bennett et al. 2002a). The orientation of
$\mathbf{\tilde{v}}$ is also affected by the fact that parallax events
are commonly due to bulge sources being lensed by foreground disk
stars (see Section \ref{sec:obs:dist}).  For the disk-bulge
configuration $\tv$ is, on average, orientated along the direction of
rotation of the Galactic plane owing to the fact that the Sun and the
lenses share a common motion in the plane of the disk. This enhances
the fraction of parallax events with
$-90^{\circ}<\theta_G<90^{\circ}$, as can be seen from
Fig.~\ref{fig:angles}.

We also investigated the radial velocities of the source stars for our
mock catalogues. No significant difference was found between the
parallax events and all lensed sources.

\subsection{Mass Estimators}
\label{sec:estimators}

Bennett et al. (2002a) and Agol et al. (2002) proposed very similar
techniques for estimating the mass of the lens for parallax
events. Using our mock catalogues, we can assess the reliability of
their estimator.

Agol et al. (2002) assume that the source and lens populations are
both characterised by Gaussian velocity distributions with means
$\langle \bmath{v_L} \rangle $ and $\langle \bmath{v_S} \rangle$ and
dispersions $\bmath{\sigma_L}$ and $\bmath{\sigma_S}$, diagonalised in
Galactic longitude and latitude ($\ell,b$). Specialising to the case
of a power-law mass function $n(m) \propto m^{-\beta}$, Agol et al.'s
(2002) maximum likelihood estimator becomes:
\begin{equation}
L(x | \bmath{ \tilde v}, \tE) \propto
x^{\beta-1}(1-x)^{5-\beta}
{\rho_{\rm L}(x) \over \sigma_\ell \sigma_b}
\exp\left[-\left(\frac{v_\ell^2}{2\sigma^2_{L,\ell}}
          + \frac{v_b^2}{2\sigma^2_{L,b}}\right)\right],
\label{eq:likelihood}
\end{equation}
where 
\begin{equation}
(v_\ell, v_b) = \langle \bmath{v_L} \rangle - x \langle \bmath{v_S}\rangle
- (1-x) ( \bmath {v_\odot} + \bmath{ \tilde v}),
\end{equation}
and
\begin{equation}
\sigma^2_{\ell} = \sigma^2_{L,\ell} + x^2 \sigma^2_{S,\ell},
\qquad \sigma^2_{b} = \sigma^2_{L,b} + x^2 \sigma^2_{S,b}.
\end{equation}
$L(x | \bmath{ \tilde v}, \tE)$ gives the likelihood of a lens lying
at a fractional distance $x = D_{\rm L}/D_{\rm S}$, given the
observables $ \bmath{\tilde v}$ and $\tE$. This is obtained using
Bayes' theorem, assuming uniform priors in lens distance and no errors on
the velocity and timescale measurements.  It can be converted to a
likelihood in mass $m$ using
\begin{equation}
m(x) = {{\tilde v}^2 \tE^2 c^2 (1-x)\over 4 G x D_{\rm S}}.
\end{equation}
As the mass functions in our simulations are more complicated
than simple power-laws, we explore two different choices, namely
$\beta = 0$ and $\beta = 1.5$. In the latter case,
equation~(\ref{eq:likelihood}) reduces to a formula proposed by
Bennett et al. (2002a).

The estimator also assumes that the distance of the source $D_{\rm S}$
is known and that the lensing population is identifiable, so that an
informed choice for the density $\rho_{\rm L}$ can be made. We
extracted 728 events from our mock catalogue that pass Wo\'zniak et
al.'s (2001) criteria. These are moderate parallax events with $\Delta
\chi^2 > 50$ and $S/N > 5$. We use the fitted values of $\bmath{
\tilde v}, \tE$, together with the known source distance $D_{\rm S}$
and the known type of deflector population to compute the estimated
mass. Of course, the accuracy in the estimate can be
calculated, as the true mass is known.

\begin{figure*}
\centering{\includegraphics[width=10cm]{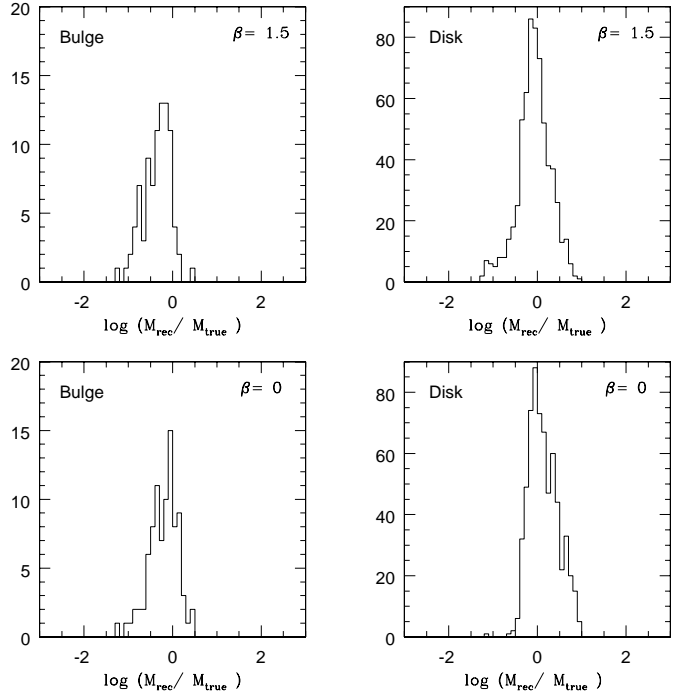}}
\caption{The error distributions of the estimators of the
mass of the lens of a parallax event using $\beta = 1.5$ (above) and
$\beta = 0$ (below) in equation~(\ref{eq:likelihood}). The results for
bulge deflectors are shown on the left, those for disk deflectors on
the right. The abscissa is the logarithm of the ratio of the recovered
to the true mass.}
\label{fig:agol}
\end{figure*}

Fig.~\ref{fig:agol} shows histograms of the logarithm of the
ratio of the recovered mass to the true mass for the 89 events caused
by bulge deflectors (left panels) and 639 events caused by disk
deflectors (right panels) using the likelihood estimator with $\beta =
1.5$ and $\beta = 0$. For the bulge lenses, both estimators give a
similar performance with mean percentage errors of $-22$ per cent
($\beta =0$) and $-45$ per cent ($\beta = 1.5$). A negative 
value of the mean percentage error implies that the estimators
typically underestimate the mass.  For the disk lenses the estimators
tend to overestimate the mass, with a mean percentage error of $85$ per
cent for $\beta =0$, compared to $20$ per cent for $\beta = 1.5$.  The
simulations use more complicated mass functions than power-laws.  It
is noteworthy that the effect of using a simple power-law mass
function in the likelihood estimator (which is needed to perform the
integrals) is to cause systematic offsets in the recovered masses as
compared to the true masses.

The interpretation of the effectiveness of these mass estimators
suffers from an additional complication. As was shown in Section
\ref{sec:def}, the accurate recovery of the parallax parameters such
as $\mathbf{\tilde v}$ is not guaranteed, even for our convincing
parallax events (see Fig. \ref{fig:paraerr}). Clearly these mass
estimators cannot be expected to perform reliably if the parallax
parameters have not been determined accurately from the observed
light curves.

\subsection{Long Duration Events and Black Hole Candidate Lenses}
\label{sec:blackhole}

Our simulations can also be used to investigate the nature of observed
long duration microlensing events, a number of which have been
speculated to be caused by black hole lenses.

In our model, we do not include any stellar remnant populations.
However, white dwarfs ($M\sim0.6~\mbox{M}_{\sun}$), neutron stars
($M\sim1.4~\mbox{M}_{\sun}$) and black holes ($M\sim
\mbox{several}~\mbox{M}_{\sun}$) must exist in the Galactic disk, and
so their contribution to parallax events is not accounted for in our
model. For long durations ($t_{\rm E}\ga100~\mbox{d}$), the time-scale
distribution asymptotically approaches a power-law (Mao \& Paczy\'nski
1996). As a result, the fraction of events contributed by stellar
remnants reaches an asymptotic value, determined entirely by the mass
function. The relative fraction is weighted according to $\sim M^2
n(M) dM$, and hence favours massive lenses (Agol et al. 2002). If we
adopt a mass function for stellar remnants (e.g., Gould 2000; Han \&
Gould 2003), we find
that the fraction of long duration events contributed by stellar
remnants is about 56 per cent. Therefore our predicted parallax
fraction may be too low by a factor of $\sim$2. However, in reality,
the discrepancy may be less than this because on average massive
stellar remnants will have larger $\rEt$ and correspondingly a reduced
probability of exhibiting parallax signatures (see Fig.
\ref{fig:pararEt} and Section~\ref{sec:rEdist}). In addition, since
the baseline of our mock light curves is only 3 yr, events with such
long time-scales may not have been considered in this analysis. For a
mock event to be included in our parallax catalogue, we require a
constant baseline, which excludes exceptionally long duration events.

To date, the most promising candidate black hole event is
OGLE-1999-BUL-32/MACHO-99-BLG-22 (Mao et al. 2002; Bennett et
al. 2002b). This event, along with an additional two events
MACHO-96-BLG-5 and MACHO-98-BLG-6 (Bennett et al. 2002a), have been
analysed by Agol et al. (2002). In this paper, they used their mass
estimator (see equation \ref{eq:likelihood}) to conclude that event
OGLE-1999-BUL-32 has a probability of 76 per cent of being caused by a
black hole lens (with the remaining two events having probabilities
less than 20 per cent). The event parameters $\tE$ and $\rEt$ for
these three events can be found in Table \ref{table:par}. In addition
to $\tE$ and $\rEt$, Agol et al. (2002) also incorporated the angle of
the projected velocity $\theta_G$ and the lower limit on the lens
magnitude into their estimator. For events OGLE-1999-BUL-32,
MACHO-96-BLG-5 and MACHO-98-BLG-6, the respective angles and $I$-band
lens magnitude constraints are: $142^{\circ}$, $I_{\rm L}>18.6$;
$-20^{\circ}$, $I_{\rm L}>18.6$; $-68^{\circ}$. The lower limit for
the $I$-band lens magnitude of event MACHO-98-BLG-6 is unknown since
Bennett et al. (2002a) only constrain the $V$-band magnitude.

\begin{figure*}
{\centering\includegraphics[width=17cm]{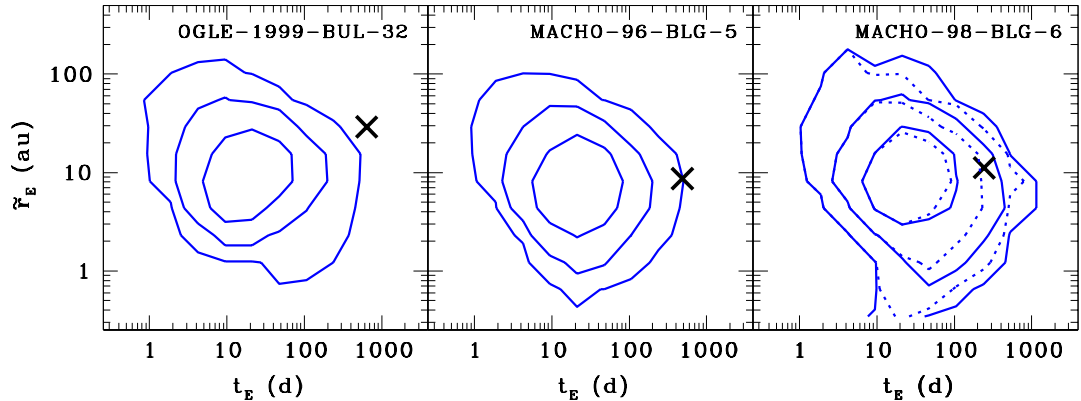}}
\caption{Contour plot showing the joint distribution of event
time-scale and projected Einstein radius on the observer plane for all
events regardless of whether or not they pass the OGLE-II detection
criteria of Wo\'zniak et al. (2001). The contours denote the 68.3,
95.4 and 99.7 per cent confidence regions. Each panel corresponds to the
mock events that have projected velocity angle $\theta_G$ and lens
magnitude consistent with the observed black hole candidate
named. Note that no $I$-band magnitude limit is available for event
MACHO-1998-BLG-6 and hence we provide two sets of contours: the solid
line shows the effect of including no restriction on the lens
magnitude and the dotted line corresponds to the limit $I_{\rm L} >
21.2$, which is obtained from the $V$-band limit $V_{\rm L} > 22.2$
(Bennett et al. 2002a) and a fiducial estimate of the colour of the lens
$(V-I)_{\rm L}=1.0$. Note that since one might na\"{\i}vely suspect
that $(V-I)_{\rm L}\ga1.0$, the true contours should lie somewhere
between the dotted and solid contours presented here.}
\label{fig:rEtEcont}
\end{figure*}

Using our simulations we now attempt to determine whether the
observed characteristics of these events are consistent with our
stellar population (that contains no remnants).  In
Fig. \ref{fig:rEtEcont}, we calculate the joint probability
distribution of the parameters $\rEt$ and $\tE$ for the three black
hole candidate events; for each event, we plot the probability
distributions for the mock events that match the observed $\theta_G$
and have lens magnitudes fainter than the required limit. Note that
here we do not incorporate the OGLE-II event detection efficiency,
since many of these long duration events will be omitted due to the
3-yr baseline of the experiment. From this figure, it can be concluded
that although the properties of event MACHO-1998-BLG-6 are consistent
with what is predicted from our simulations, the remaining two events,
especially OGLE-1999-BUL-32, are clearly inconsistent. However, such
large values of $\rEt$ can be obtained by having the lens and source
close together, such as from bulge self-lensing. Although bulge
self-lensing events typically have a shorter time-scale (with a mean
$\tE$ of 26.3 d versus 32.1 d for all events), if the Galactic model
incorporated the streaming motion of the bar then it may be feasible
to obtain both large $\rEt$ and $\tE$ simultaneously.

\section{Discussion and Conclusions}
\label{sec:disc}

This paper has investigated the properties of parallax microlensing
events towards the Galactic bulge. The expected fraction of parallax
events based on standard models of the Galaxy is of the order of a few
per cent, which is compatible with most observed findings.
However, it should be noted that observed catalogues feature other
types of exotic microlensing events, such as binary-lens events (see
Jaroszy\'{n}ski 2002) and binary-source (xallarap) events, many of
which can mimic parallax signatures and hence
display an improvement in $\chi^2$ when fit with a parallax model
(e.g. Smith et al. 2002b). Although our predicted parallax fractions
seem consistent with observed catalogues, there is one exception,
namely the results from the EROS collaboration (Afonso et
al. 2003). In this paper, they 
report a fraction approximately 12.5 per cent, which is even more
surprising when one considers that this catalogue is from clump-giant
sources, which supposedly reside in the bulge; our simulations predict
that the fraction of parallax events for bulge sources is only $\sim1$
per cent. It could be argued that the EROS source stars are bright red
clump giant stars and as such may have higher S/N than our
simulations. However, even if we restrict our simulations to bright
bulge source stars with high S/N, the percentage of events with
convincing parallax signatures is only 2 per cent; given this
predicted fraction, the probability that such a sample of 16 events
will yield two parallax events is only 0.04. 
Therefore, although no strong conclusions can be made, it seems that this
observed fraction of parallax events from the EROS catalogue may be
inconsistent with our simulations.

Our models have one major shortcoming, namely that we only
included blending from the lenses themselves, i.e. we did not include
blending from nearby unrelated stars. We attempted to limit this
problem by only considering sources brighter than $I=19$, since it is
commonly assumed that such bright sources are less affected by
blending.\footnote{However, recent work by Sumi et al. (2005) has
suggested that even this assumption may not be wholly reliable.} To
test what effect blending from unrelated stars would have on our
simulations, we perform the following simple investigation; we generate
an additional catalogue of mock events with blending ratio (i.e. ratio
of source flux to total baseline flux) distributed uniformly between
zero and one. This means that the source stars for the blended
catalogue are fainter than the corresponding source star in the
unblended catalogue. 
Although the mean time-scale of observed events ($t_{\rm E}$) is
practically unchanged in this new catalogue, the overall fraction of 
parallax events is slightly reduced due to the greater number of lower
signal-to-noise ratio events, which can be expected since the source
stars are fainter for the blended catalogue. Currently, it is unclear
how blending varies as a function of source magnitude and event
time-scales, so we cannot address this question
quantitatively. However, the qualitative differences between all
microlensing and parallax events (i.e. the distributions shown in
Section \ref{sec:obs}) appear unchanged from this analysis of blending.

\begin{figure}
\plotone{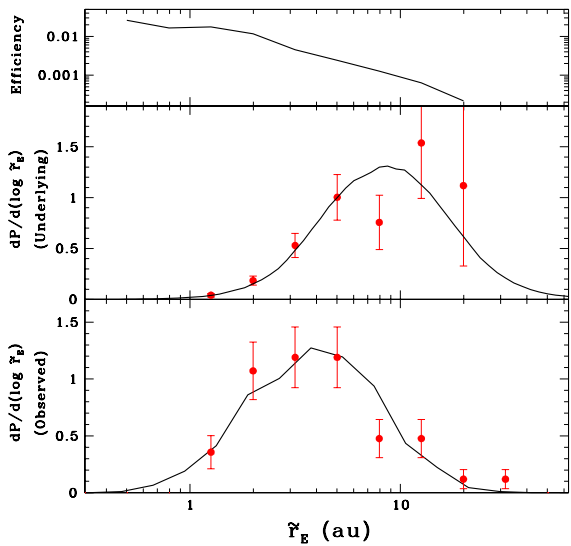}
\caption{A comparison between the distribution of the projected
Einstein radius ($\rEt$) for observed parallax events and our mock
catalogues. The distributions from our mock catalogue
are given by the solid lines and the observed distributions are given
by the data points (where Poisson errors have been assumed). The lower
panel shows the observed distribution, while the middle panel shows
the underlying distribution, after correcting for the efficiency (top
panel).}
\label{fig:rEtobs}
\end{figure}

To obtain a greater understanding of the underlying nature of parallax
events, we examined the distributions of the event parameters $t_{\rm
E}$ and $\tilde v$ (or, equivalently, $\rEt$) for parallax events,
comparing them to the distributions for all events. One
important question is how well do our simulations match the observed
characteristics of parallax events. Although the overall fractions of
parallax events seem consistent with current catalogues, we can also
tentatively compare the parallax parameters such as the projected
Einstein radius, $\rEt$. In Fig. \ref{fig:rEtobs}, we show a plot
comparing the predictions from our simulations with the distribution
of $\rEt$ for the observed parallax events from Table
\ref{table:par}. It should be noted that since the events from Table
\ref{table:par} come from a wide range of microlensing experiments
with differing durations, sampling and photometric properties, any
comparison can only be very speculative. However, as can be seen from
the bottom panel of this figure, the observed distribution of $\rEt$
(as given by the data points) appears to be in reasonable agreement
with the distribution from parallax events in our mock
catalogues. Furthermore, from our simulations, we can calculate the
efficiency of recovering parallax signatures and attempt to
determine the form of the underlying $\rEt$ distribution for these
observed events. This method is analogous to the one used to convert
an observed time-scale distribution into a true underlying time-scale
distribution using the `detection efficiency'. The upper panel of
Fig. \ref{fig:rEtobs} shows the efficiency of detecting parallax
signatures, and when applied to the observed $\rEt$ distribution we
obtain the underlying distribution shown in the middle panel. It
appears that the observed distribution is consistent with the
underlying $\rEt$ distribution from our simulations.

Although there are currently too few observed parallax events for
us to make any firm statements about the distribution of the parallax
parameters, it may become possible in the near future for
projects such as OGLE-III (Udalski 2003), which is currently detecting
$\sim$500 microlensing events each year. 
Assuming the rate of parallax
detection for this project is similar to that of OGLE-II, the
three-fold increase in sky coverage means that a few good quality
parallax events will be detected each year.
It is expected that the OGLE-III project (or an upgraded experiment,
OGLE-IV) will continue for many years, which should be sufficient to
enable the construction of distributions of $\rEt$ and $\tilde v$ that
are reasonably well-constrained. A similar study to the
present one would allow a detailed comparison between theoretical
predictions and observations for parallax events.

Unique lens mass determinations for parallax events may become routine
in the near future by combining a reliable measurement of $\rEt$ with
additional constraints, such as has been shown already for a number of
events (An et al. 2002; Gould et al. 2004; Kubas et al. 2005). One
particular approach that has been proposed is the measurement of the
separation between the two microlensed images using powerful
interferometers such as VLTI (Delplancke, G\'orski \& Richichi
2001). We can see from our simulations that such image separations at
maximum magnification ($\theta_{\rm sep}=\theta_{\rm E}\sqrt{u_0^2+4}$,
where $u_0$ is the impact parameter corresponding to the maximum
magnification) are larger for our parallax sample, with mean
separations of 1.8 mas compared to 0.8 mas for the full
sample. Therefore it will be easier to resolve the two microlensed
images for parallax events and, in addition, their longer duration
will make it more feasible to plan high signal-to-noise ratio
observations while the events are still undergoing high magnification.

\section*{acknowledgments}

MCS and VB acknowledge financial support by PPARC. MCS also
acknowledges financial support from the Netherlands Organisation for
Scientific Research (NWO). This work has benefited from the financial
help from the visitor's grants at Jodrell Bank and Cambridge.
This work was partially supported by the European Community's Sixth
Framework Marie Curie Research Training Network Programme, Contract
No. MRTN-CT-2004-505183 `ANGLES'. We are deeply indebted to the
referee Andy Gould for providing a comprehensive and insightful
report. We also wish to thank Pavel Kroupa for advice regarding the
mass and luminosity functions.

\end{document}